\documentclass{cernrep}
\usepackage{axodraw}

\begin{document}

\begin{flushright}
hep-ph/0611247\\
CERN-LCGAPP-2006-06\\
November 2006\\[10mm]
\end{flushright}

\title{Monte Carlo Generators%
\footnote{Lectures presented at the 2006 European School of
High-Energy Physics, Aronsborg, Sweden, 18 June -- 1 July 2006}}
\author{Torbj\"orn Sj\"ostrand}
\institute{CERN/PH, CH--1211 Geneva 23, Switzerland, and\\
Department of Theoretical Physics, Lund University,
S\"olvegatan 14A, SE--223 62 Lund, Sweden}
\maketitle

\begin{abstract}
The structure of events in high-energy collisions is complex and not
predictable from first principles. Event generators allow the problem
to be subdivided into more manageable pieces, some of which can be
described from first principles, while others need to be based on 
appropriate models with parameters tuned to data. In these lectures 
we provide an overview, discuss how matrix elements are used, 
introduce the machinery for initial- and final-state parton showers, 
explain how matrix elements and parton showers can be combined for
optimal accuracy, introduce the concept of multiple parton--parton
interactions, comment briefly on the hadronization issue, and provide 
an outlook for the future.
\end{abstract}

\section{Introduction}

Given the current landscape in experimental high-energy physics, 
these lectures are focused on applications of event generators 
for hadron colliders like the Tevatron and LHC. Much of it would
also be relevant for $e^+e^-$ machines like LEP and ILC or $e^{\pm}p$ 
machines like HERA, but with some differences not discussed here. 
Heavy-ion physics is not at all addressed, since it involves rather 
different aspects, specifically the potential formation of a 
quark--gluon plasma. Further, within the field of high-energy 
$pp/p\overline{p}$ collisions, the emphasis will be on the common 
aspects of QCD physics that occurs in all collisions, rather on those 
aspect that are specific to a particular physics topic, such as $B$ 
production or supersymmetry. Both heavy ions and other physics topics 
are instead covered by other lectures at this school.

Section 2 contains a first overview of the physics picture and
the generator landscape. Thereafter section 3 describes the usage of
matrix elements, section 4 the important topics of initial- and 
final-state showers, and section 5 how showers can be matched
to different hard processes. The issue of multiple interactions and 
their role in mimimum-bias and underlying-event physics is introduced
in section 6, followed by some brief comments on hadronization in 
section 7. The article concludes with an outlook on the ongoing 
generator-development work in section 8. 

Slides for these and other similar lectures \cite{otherlectures} 
are complementary to this writeup in style and contents, 
including many (colour) illustrations absent here.
Other useful resources include the ``Les Houches Guidebook to Monte Carlo 
Generators for Hadron Collider Physics'' \cite{guidebook} and a recent
review on QCD physics at the Tevatron and LHC \cite{chs}.
 
\section{Overview}

In real life, machines produce events that are stored by the data
acquisition system of a detector. In the virtual reality, event 
generators like \textsc{Herwig} \cite{herwigsix} and \textsc{Pythia} 
\cite{pythiasix} play the role of machines like the Tevatron and LHC, 
and detector simulation programs like \textsc{Geant~4} \cite{geant} 
the role of detectors like ATLAS or CMS. The real and virtual worlds 
can share the same event reconstruction framework and subsequent physics 
analysis. It is by understanding how an original physics input is 
distorted step-by-step in the better-controlled virtual world that   
an understanding can be gained of what may be going on in the real world.
For approximate studies the detector simulation and reconstruction 
steps can be shortcut, so that generators can be used directly 
in the physics studies.

A number of physics analyses would not be feasible without generators.
Specifically, a proper understanding of the (potential) signal and 
background processes is important to separate the two. The key aspect 
of generators here is that they provide a detailed description of
the final state so that, ideally, any experimental observable or
combination of observables can be predicted and compared with data.
Thereby generators can be used at various stages of an experiment:
when optimizing the detector and its trigger design to the intended 
physics program, when estimating the feasibility of a specific physics 
study, when devising analysis strategies, when evaluating acceptance 
corrections, and so on. 

However, it should always be kept in mind that generators are not perfect. 
They suffer from having to describe a broad range of physics, some of 
which is known from first principles, while other parts are modelled in 
different frameworks. (In the latter case, a generator actually acts as a 
vehicle of ideology, where ideas are disseminated in prepackaged form from 
theorists to experimentalists.) Given the limited resources, different 
authors may also have invested more or less time on specific physics 
topics, and therefore these may be more or less well modelled. It
always pays to shop around, and to compare several approaches before
drawing too definite conclusions. Blind usage of a generator is not
encouraged: then you are the slave rather than the master.   

Why then \textit{Monte Carlo} event generators? Basically because 
Einstein was wrong: God does throw dice! In quantum meachanics, 
calculations provide the \textit{probability} for different outcomes 
of a measurement. Event-by-event, it is impossible to know beforehand 
what will happen: anything that is at all allowed could be next.
It is only when averaging over large event samples that the expected 
probability distributions emerge --- provided we did the right 
calculation to high enough accuracy. In generators, (pseuo)random numbers 
are used to make choices intended to reproduce the quantum mechanical 
probabilities for different outcomes at various stages of the process.  

The buildup of the structure in an event occurs in several steps, and
can be summarized as follows:
\begin{itemize}
\item Initially two hadrons are coming in on a collision course.
Each hadron can be viewed as a bag of partons --- quarks and gluons.
\item A collision between two partons, one from each side, gives the
hard process of interest, be it for physics within or beyond the 
standard model: $u g \to u g$, $u \overline{d} \to W^+$, $g g \to h^0$,
etc. (Actually, the bulk of the cross section results 
in rather mundane events, with at most rather soft jets, or of a 
simple elastic or diffractive character that is not easily described 
as partonic processes. Such events usually are filtered away at an early
stage, however.)
\item When short-lived ``resonances'' are produced in the hard process, 
such as the top, $W^{\pm}$ or $Z^0$, their decay has to be viewed as part 
of this process itself, since e.g. spin correlations are transferred
from the production to the decay stages.
\item A collision implies accelerated colour (and often electromagnetic)
charges, and thereby brems\-strah\-lung can occur. Emissions that can be 
associated with the two incoming colliding partons are called 
Initial-State Radiation (ISR). As we shall see, such emissions can  
be modelled by so-called space-like parton showers.
\item Emissions that can be associated with outgoing partons are instead
called Final-State Radiation (FSR), and can be approximated be
time-like parton showers. Often the distinction between a hard process and
ISR and FSR is ambiguous, as we shall see. 
\item So far we only extracted one parton from each incoming hadron to
undergo a hard collision. But the hadron is made up of a multitude of 
further partons, and so further parton pairs may collide within one
single hadron--hadron collision --- multiple interactions (MI). (Not to be 
confused with pileup events, when several hadron pairs collide during a
bunch--bunch crossing, but with obvious analogies.) 
\item Each of these further collisions also may be associated with its 
ISR and FSR.
\item The colliding partons take a fraction of the energy of the incoming
hadrons, but much of the energy remains in the beam remnants, which
continue to travel essentially in the original directions. These 
remnants also carry colours that compensate the colour taken away by
the colliding partons.
\item At short timescales, when partons are close to each other, the 
principle of asymptotic freedom tells us that we can think of each parton 
as freely moving along its trajectory. However, as time goes by and the 
partons recede from each other, confinement forces become significant.
The structure and time evolution of these force fields cannot be described 
from first principles within any calculational technique currently at our 
disposal, so models have to be introduced. One common approach is to 
assume that a separate confinement field is stretched between each colour 
and its matching anticolour, with each gluon considered as a simple sum 
of a colour and an anticolour, and all colours distinguishable from each 
other (the $N_C \to \infty$ limit). 
\item Such fields can break up by the production of new quark--antiquark
pairs that screen the endpoint colours, and where a quark from one break
(or from an endpoint) can combine with an antiquark from an adjacent 
break to produce a primary hadron. This process is called hadronization.
\item Many of those primary hadrons are unstable and decay further at 
various timescales. Some are sufficiently long--lived that their decays
are visible in a detector, or are (almost) stable. Thereby we have reached 
scales where the event-generator description has to be matched to a 
detector-simulation framework. 
\item It is only at this stage that experimental information can
be obtained and used to reconstruct back what may have happened at the 
core of the process, e.g.\ whether a Higgs particle was produced or not.
\end{itemize}

The Monte Carlo method allows these steps to be considered sequentially,
and within each step to define a set of rules that can be used 
iteratively to construct a more and more complex state, maybe ending
with hundreds of particles moving out in different directions. Since
each particle contains of the order of ten degrees of freedom 
(flavour, mass, momentum, production vertex, lifetime, \ldots) we 
realize that thousands of choices are involved for a typical event.
The aim is to have a sufficiently realistic description of these choices
that both the average behaviour and the fluctuations around this average
are well decribed. 

Schematically, the cross section for a range of final states is provided by
\begin{equation}
\sigma_{\mathrm{final~state}} = \sigma_{\mathrm{hard~process}} \, 
\mathcal{P}_{\mathrm{tot},\mathrm{hard~process} \to \mathrm{final~state}}
~,
\end{equation}
properly integrated over the relevant phase-space regions and summed
over possible ``paths'' (of showering, hadronization, etc.) that lead 
from a hard process to the final state. That is, the dimensional 
quantities are associated with the hard process; subsequent steps are 
handled in a probabilistic approach.

The spectrum of event generators is very broad, from general-purpose ones 
to more specialized ones. \textsc{Herwig} and \textsc{Pythia} are the
two most commonly used among the former ones, with \textsc{Isajet}
\cite{isajet} and \textsc{Sherpa} \cite{sherpa} as the other two main 
programs in this category. Among more specialized programs, many deal 
with the matrix elements for some specific set of processes, a few with 
topics such as parton showers or particle decays, but there are e.g.\ no 
freestanding programs that handle hadronization. In the end, many of the 
specialized programs are therefore used as ``plugins'' for the 
general-purpose ones.
 
\section{Matrix elements and their usage}
 
{}From the Lagrangian of a theory the Feynman rules can be derived, 
and from them matrix elements can be calculated. Combined with 
phase space it allows the calculation of cross sections. As a simple
example consider the scattering of quarks in QCD, say 
$u(1) \, d(2) \to u(3) \, d(4)$, a process similar to Rutherford 
scattering but with gluon exchange instead of photon ditto. 
The Mandelstam variables are defined 
as $\hat{s} = (p_1 + p_2)^2$, $\hat{t} = (p_1 - p_3)^2$ and  
$\hat{u} = (p_1 - p_4)^2$. In the cm frame of the collision $\hat{s}$
is the squared total energy and 
$\hat{t}, \hat{u} = - \hat{s} (1 \mp \cos\hat{\theta})/2$ where
$ \hat{\theta}$ is the scattering angle. The differential cross section
is then
\begin{equation}
\frac{\mathrm{d} \hat{\sigma}}{\mathrm{d} \hat{t}} = 
\frac{\pi}{\hat{s}^2} \, \frac{4}{9} \, \alpha_{\mathrm{s}}^2 \, 
\frac{\hat{s}^2 + \hat{u}^2}{\hat{t}^2} ~,
\end{equation} 
which diverges roughly like $\mathrm{d}p_{\perp}^2/p_{\perp}^4$
for transverse momentum $p_{\perp} \to 0$. We will come back to this 
issue when discussing multiple interactions; for now suffice to say that
some lower cutoff $p_{\perp\mathrm{min}}$ need to be introduced. 
Similar cross sections, differing mainly by colour factors, are 
obtained for $q \, g \to q \, g$ and $g \, g \to g \, g$. A few further
QCD graphs, like $g \, g \to q \, \overline{q}$, are less singular
and give smaller contributions.
These cross sections then have to be convoluted with the flux of the 
incoming partons $i$ and $j$ in the two incoming hadrons $A$ and $B$:
\begin{equation}
\sigma = \sum_{i,j} \int \!\!\! \int \!\!\! \int 
\mathrm{d} x_1 \,  \mathrm{d} x_2 \, \mathrm{d} \hat{t} \, 
f_i^{(A)}(x_1, Q^2) \, f_j^{(B)}(x_2, Q^2) \, 
\frac{\mathrm{d} \hat{\sigma}_{ij}}{\mathrm{d} \hat{t}}  ~.
\label{eq:sigma}
\end{equation} 
The parton density functions (PDFs) of gluons and sea quarks are 
strongly peaked at small momentum fractions $x_1 \approx E_i/E_A$, 
$x_2 \approx E_j/E_B$. This further enhances the peaking of the 
cross section at small $p_{\perp}$ values. Nevertheless, with high 
machine luminosity the jet cross section can be studied out to quite 
high values. 

The cross section of other processes can be suppressed by two main effects. 
Firstly, for massive particles the $p_{\perp}$ spectrum is strongly 
dampened below the respective mass scale, and it is only above it 
that these precesses have a chance to stand up above the QCD  background. 
Secondly, the processes may involve electroweak (or other small) 
couplings rather than strong ones. 

In order to address the physics of interest a large number of processes, 
both within the Standard Model and in various extensions of it, have to 
be available in generators. Indeed many can also be found in the 
general-purpose ones, but by far not enough. Further, often processes 
are there available only to lowest order, while experimental interest
 may be in higher orders, with more jets in the final state, either as 
a signal or as a potential background. So a wide spectrum of 
matrix-element-centered programs are available \cite{hepcode},
some quite specialized and others more generic. 
  
\begin{figure}[t]  
\begin{center}    
\begin{picture}(320,210)(0,0)
\BBox(100,185)(220,205)
\Text(160,195)[]{Process Selection}
\BBox(100,165)(220,185)
\Text(160,175)[]{Resonance Decays}
\LongArrow(160,165)(160,152)
\BBox(100,130)(220,150)
\Text(160,140)[]{Parton Showers}
\BBox(100,110)(220,130)
\Text(160,120)[]{Multiple Interactions}
\BBox(100,90)(220,110)
\Text(160,100)[]{Beam Remnants}
\LongArrow(160,90)(160,77)
\BBox(100,55)(220,75)
\Text(160,65)[]{Hadronization}
\BBox(100,35)(220,55)
\Text(160,45)[]{Ordinary Decays}
\LongArrow(160,35)(160,22)
\BBox(100,0)(220,20)\Text(160,10)[]{Detector Simulation}
\BBox(0,185)(80,205)
\Text(40,195)[]{ME Generator}
\LongArrow(40,185)(40,172)
\BBox(0,150)(80,170)
\Text(40,160)[]{ME Expression}
\LongArrow(80,165)(98,195)
\LongArrow(80,160)(98,180)
\BBox(0,85)(80,135)
\Text(40,125)[]{SUSY/\ldots}
\Text(40,110)[]{spectrum}
\Text(40,96)[]{calculation}
\LongArrow(40,135)(40,148)
\LongArrow(80,125)(98,170)
\BBox(240,170)(320,205)
\Text(280,195)[]{Phase Space}
\Text(280,180)[]{Generation}
\LongArrow(230,195)(238,190)
\LongArrow(230,195)(222,200)
\BBox(240,115)(320,135)
\Text(280,125)[]{PDF Library}
\LongArrow(240,130)(222,190)
\LongArrow(240,125)(222,140)
\LongArrow(240,120)(222,120)
\BBox(240,60)(320,80)
\Text(280,70)[]{$\tau$ Decays}
\LongArrow(240,70)(222,50)
\BBox(240,25)(320,45)
\Text(280,35)[]{$B$ Decays}
\LongArrow(240,35)(222,40)
\DashArrowArcn(100,28.75)(13.75,270,90){5}
\DashArrowArcn(100,33.75)(28.75,270,90){5}
\SetWidth{1}
\LongArrow(80,155)(98,120)
\end{picture}
\end{center}
\caption{Example how different programs can be combined in the
event-generation chain. \label{fig:combgen}}
\end{figure}
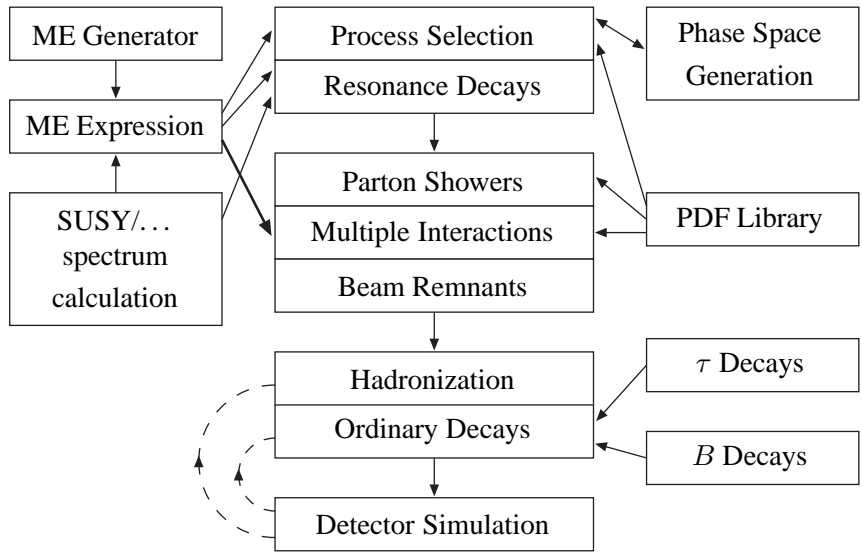

The way these programs can be combined with a general-purpose generator
is illustrated in Fig.~\ref{fig:combgen}. In the study of Supersymmetry
(SUSY) it is customary to define a model in terms of a handful parameters, 
e.g.\ specified at some large Grand Unification scale. It is then the task 
of a spectrum calculator to turn this into a set of masses, mixings and
couplings for the physical states to be searched for. Separately, the 
matrix elements can be calculated with these properties as unknown 
parameters, and only when the two are combined is it possible to speak
of physically relevant matrix-element expressions. These matrix 
elements now need to be combined with PDF's and sampled in phase space,
preferable with some preweighting procedure so that regions of 
phase space with high cross sections are sampled more frequently.
The primarily-produced SUSY particles typically are unstable and undergo
sequential decays down to a lightest supersymmetric particle (LSP), 
again with branching ratios and angular distributions that should be 
properly modelled. The LSP would be neutral and escape undetected,
while other decay products would be normal quarks and leptons. 

It is at this stage that general-purpose programs take over. They 
describe the showering associated with the above process, the presence 
of additional interactions in the same hadron--hadron collision, 
the structure of beam remnants, and the hadronization and decays. 
They would still rely on the externally supplied PDF's, and potentially
make use of programs dedicated to $\tau$ and $B$ decays, where 
spin information and form factors require special encoding. Even after
the event has been handed on to the detector-simulation program 
some parts of the generator may be used in the simulation of secondary
interactions and decays.

Several standards have been developed to further this interoperability.
The Les Houches Accord (LHA) for user processes \cite{lha} 
specifies how parton-level information about the hard process and 
sequential decays can be encoded and passed on to a general-purpose 
generator. Originally it was defined in terms of two Fortran commonblocks, 
but more recently a standard Les Houches Event File format \cite{lhef} 
offfers a language-independent alternative approach. The Les Houches 
Accord Parton Density Functions (LHAPDF) library \cite{lhapdf} makes
different PDF sets available in a uniform framework. The SUSY Les Houches 
Accord (SLHA) \cite{slha} allows a standardized transfer of masses, mixings, 
couplings and branching ratios from spectrum calculators to other programs.
Finally, the HepMC C++ event record \cite{hepmc} succeeds the 
HEPEVT Fortran one \cite{hepevt} as a standard way to transfer information
from a generator on to the detector-simulation stage. One of the key
building blocks for several of these standards is the PDG codes for 
all the most common particles \cite{pdg}, also in some scenarios for 
physics beyond the Standard Model.    

The $2 \to 2$ processes we started out with above are about the simplest 
one can imagine at a hadron collider. In reality one needs to go on to 
higher orders. In $\mathcal{O}(\alpha_{\mathrm{s}}^3)$ two new kind of 
graphs enter. One kind is where one additional parton is present in the final 
state, i.e.\ $2 \to 3$ processes. The cross section for such processes is 
almost always divergent when one of the parton energies vanish (soft 
singularities) or two partons become collinear (collinear singularities). 
The other kind is loop graphs, with an additional intermediate parton not 
present in the final state, i.e.\ a correction to the $2 \to 2$ processes.     
Strictly speaking, at $\mathcal{O}(\alpha_{\mathrm{s}}^3)$ one picks up the 
interference between the lowest-order graph and the loop graph, and this 
interference has negative divergences that exactly cancel the positive ones 
above, with only finite terms surviving. For inclusive event properties 
such next-to-leading order (NLO) calculations lead to an improved accuracy 
of predictions, but for more exclusive studies the mathematical cancellation
of singularities has to be supplemented by more physical techniques,
which is far from trivial. 

The tricky part of the calculations is the virtual corrections. NLO is
now state-of-the-art, with NNLO still in its infancy. If one is content with 
Born-level diagrams only, i.e.\ without any loops, it is possible to go to 
quite high orders, with up to something like eight partons in the final state. 
These partons have to be kept well separated to avoid the phase-space 
regions where the divergences become troublesome. In order to cover also 
regions where partons become soft/colliner we therefore next turn our 
attention to parton showers. 
 
\section{Parton showers}

To iterate, the emission rate for a branching such as $q \to q g$ diverges 
when the gluon either becomes collinear with the quark or when the gluon 
energy vanishes. The QCD pattern is similar to that for $e \to e \gamma$ 
in QED, except with a larger coupling, and actually a coupling that also 
increases for smaller relative $p_{\perp}$ in a branching, thereby further 
enhancing the divergence. Furthermore the non-Abelian character of QCD 
leads to $g \to g g$ branchings with similar divergences, without any 
correspondence in QED. The third main branching, $g \to q \overline{q}$ 
with its $\gamma \to e^+ e^-$ QED equivalence, does not have the soft
divergence and is less important.
 
Now, if the rate for one emission of a gluon is big, then also the rate 
for two or more will be big, and thus the need for high orders and many 
loops in matrix-element-based descriptions. With showers we introduce
two new concepts that make like easier,\\ 
(1) an iterative structure that allows simple expressions for 
$q \to q g$,  $g \to g g$ and $g \to q \overline{q}$ branchings 
to be combined to build up complex multiparton final states, and\\ 
(2) a Sudakov factor that offers a physical way to
handle the cancellation between real and virtual divergences.\\ 
Neither of the simplifications is exact, but together they allow 
us to provide sensible approximate answers for the structure of 
emissions in soft and collinear regions of phase space.

\subsection{The shower approach}

\begin{figure}[t]  
\begin{center}    
\begin{picture}(240,160)(0,-10)
\Line(20,20)(100,60)\Text(15,20)[r]{$d$}
\Line(20,140)(100,100)\Text(15,140)[r]{$u$}
\Gluon(100,60)(100,100){4}{5}
\Line(100,60)(180,20)\Text(185,20)[l]{$d$}
\Line(100,100)(180,140)\Text(185,140)[l]{$u$}
\DashCArc(100,80)(35,0,355){5}
\Text(115,80)[]{$Q^2$}
\Text(100,-10)[]{$2 \to 2$}
\Gluon(50,35)(100,20){4}{6}
\LongArrow(50,50)(68,45)\Text(45,50)[r]{$Q_2^2$}
\Gluon(50,125)(100,140){4}{6}
\LongArrow(50,110)(68,115)\Text(45,110)[r]{$Q_1^2$}
\Text(50,-10)[]{ISR}
\Gluon(140,40)(190,55){4}{6}
\LongArrow(130,27)(130,43)\Text(130,22)[t]{$Q_4^2$}
\Gluon(140,120)(190,105){4}{6}
\LongArrow(130,133)(130,117)\Text(130,138)[b]{$Q_3^2$}
\Text(150,-10)[]{FSR}
\end{picture}
\end{center}
\caption{The ``factorization'' of a $2 \to n$ process. 
\label{fig:factorize}}
\end{figure}
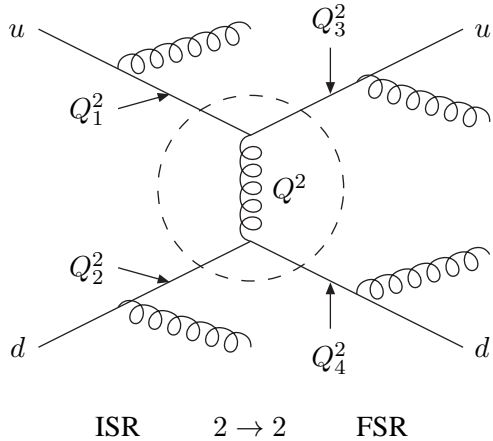

The starting point is to ``factorize'' a complex $2 \to n$ process, 
where $n$ represents a large number of partons in the final state, 
into a simple core process, maybe $2 \to 2$, convoluted with showers,
Fig.~\ref{fig:factorize}. To begin with, in a simple $u d \to u d$ process 
the incoming and outgoing quarks must be on the mass shell, i.e. satisfy
$p^2 = E^2 - \mathbf{p}^2 = m_q^2 \sim 0$, at long timescales.
By the uncertainty principle, however, the closer one comes to the 
hard interaction, i.e\ the shorter the timescales considered, the
more off-shell the partons may be. (In the oldfashioned perturbative
language particles were always on mass-shell, and the uncertainty
relation allowed energy not to be conserved temporarily. In the 
modern Feynman-graph language four-momentum is conserved at each
vertex, but intermediate ``propagator'' particles need not be on
the mass shell. The final physics is the same in both languages.) 

Thus the incoming quarks may radiate a succession of harder and harder 
gluons, while the outgoing ones radiate softer and softer gluons. 
One definition of hardness is how off-shell the quarks are, 
$ | p^2 | = | E^2 - \mathbf{p}^2 | $, 
but we will encounter other variants later. In the initial-state 
radiation (ISR) part of the cascade these virtualities are spacelike, 
$p^2 < 0$, hence the alternative name spacelike showers. 
Correspondingly the final-state radiation (FSR) is characterized by 
timelike virtualities, $p^2 > 0$, and hence also called timelike showers.

To see where this distinction comes from, consider the kinematics of an 
arbitrary branching $a \to b c$, with $a$ defined to be moving along 
the $+z$ axis. Then it is useful to introduce lightcone momenta 
$p_{\pm} = E \pm p_z$, so that the relation 
$p^2 = E^2 - p_x^2 - p_y^2 - p_z^2 = m^2$ 
translates to $p_+ p_- = m^2 + p_x^2 + p_y^2 = m^2 + p_{\perp}^2$. 
Now define the splitting of $p_+$ by $p_{+ b} = z p_{+ a}$ and
$p_{+ c} = (1 - z) p_{+ a}$. Obviously 
$\mathbf{p}_{\perp c} = - \mathbf{p}_{\perp b}$ so
$p_{\perp}^2 = p_{\perp b}^2 = p_{\perp c}^2$. Remains to ensure
conservation of $p_- = (m^2 + p_{\perp}^2) / p_+$:
\begin{equation}   
p_{- a} = p_{- b} + p_{- c} ~~\Longleftrightarrow~~ 
\frac{m_a^2}{p_{+ a}} = \frac{m_b^2 + p_{\perp}^2}{z p_{+ a}} 
 + \frac{m_c^2 + p_{\perp}^2}{(1 - z) p_{+ a}} 
~~\Longleftrightarrow~~ m_a^2 = \frac{m_b^2}{z} 
+ \frac{m_c^2}{1 - z} + \frac{p_{\perp}^2}{z (1 - z)} ~.
\label{eq:ptkin}
\end{equation} 
In an initial-state branching the incoming $a$ should be (essentially) 
massless, and if $c$ does not interact any further it should also be 
massless. This gives $m_b^2 = - (1 - z) p_{\perp}^2 < 0$, a virtuality 
that is acceptable if $b$ is on its way in to a hard scattering, i.e.\
is destined only to live for a short while. In ISR the $Q_i^2$ 
virtualities, such as $Q_1^2$ and $Q_2^2$ in Fig.~\ref{fig:factorize}, 
are usually defined as $-m_i^2$ to keep them positive definite. For a 
final-state branching, assume that $b$ and $c$ will not branch any 
further and thus are massless, while $a$ is an intermediate particle 
coming from the hard interaction. Then 
$m_a^2 = p_{\perp}^2/(z (1 - z)) > 0$ , cf.\ $Q_3^2$ and $Q_4^2$ in 
Fig.~\ref{fig:factorize}, here with $Q_i^2 = +m_i^2$. 

The cross section for the whole $2 \to n$ graph is associated with
the cross section of the hard subprocess, with the approximation that
the other $Q_i^2$ virtualities can be neglected in the matrix-element
expression. In the limit that all the $Q_i^2 \ll Q^2$ this should be 
a good approximation. In other words, first the hard process can be
picked without any reference to showers, and only thereafter are
showers added with unit probability. But, of course, the showers 
do modify the event shape, so at the end of the day the cross section 
is affected. For instance, the total transverse energy 
$E_{\perp\mathrm{tot}}$ of an event is increased by ISR, so the 
cross sections of events with a given $E_{\perp\mathrm{tot}}$
is increased by the influx of events that started out with a lower
$E_{\perp\mathrm{tot}}$ in the hard process.
 
It is important that the hard-process scale $Q^2$ is picked to be the 
largest one, i.e. $Q^2 > Q_i^2$ in Fig.~\ref{fig:factorize}. If e.g.\ 
$Q_1^2 > Q^2$ then instead the $u g \to u g$ subgraph ought to be chosen
as hard process, and the gluon of virtuality $Q^2$ ought to be part 
of the ISR off the incoming $d$. Without such a criterion one might
doublecount a given graph, or even count it once for every possible 
subgraph inside the complete $2 \to n$ graph. In addition, the 
approximation of neglecting virtualities in the hard-scattering 
matrix elements obviously becomes worse the more the incoming and 
outgoing partons are off-shell, another reason not to put a larger
scale than necessary in the shower part.  

\subsection{Final-state radiation}

\begin{figure}[t]  
\begin{center}    
\begin{picture}(180,90)(0,-50)
\Photon(0,0)(40,0){4}{4}\Text(20,15)[]{0}
\ArrowLine(40,0)(110,-40)\Text(120,-40)[l]{$q$ (1)}
\ArrowLine(110,40)(40,0)\Text(120,40)[l]{$\overline{q}$ (2)}
\Text(55,-18)[]{$i$}
\Gluon(75,-20)(110,-10){4}{4}\Text(120,-10)[l]{$g$ (3)}
\Text(70,-50)[]{(a)}
\end{picture}
\begin{picture}(150,90)(0,-50)
\Photon(0,0)(40,0){4}{4}\Text(20,15)[]{0}
\ArrowLine(40,0)(110,-40)\Text(120,-40)[l]{$q$ (1)}
\ArrowLine(110,40)(40,0)\Text(120,40)[l]{$\overline{q}$ (2)}
\Text(55,18)[]{$i$}
\Gluon(75,20)(110,10){4}{4}\Text(120,10)[l]{$g$ (3)}
\Text(70,-50)[]{(b)}
\end{picture}
\end{center}
\caption{The two Feynman graphs that contribute to 
$\gamma^*/Z^0(0) \to q(1) \, \overline{q}(2) \, g(3)$
\label{fig:qqg}}
\end{figure}
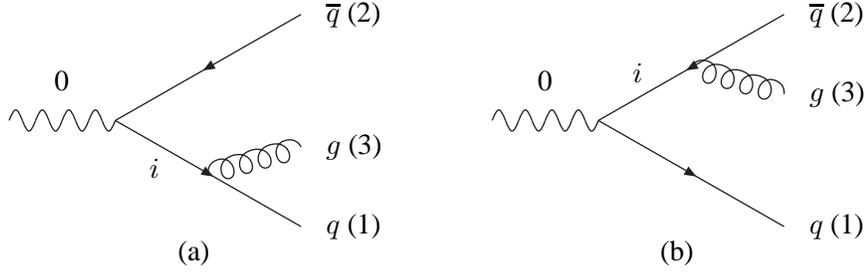

Let us next turn to a more detailed presentation of the showering
approach, and begin with the simpler final-state stage. This is most
cleanly studied in the process 
$e^+ e^- \to \gamma^*/Z^0 \to q \overline{q}$. The first-order 
correction here corresponds to the emission of one additional gluon,
by either of the two Feynman graphs in Fig.~\ref{fig:qqg}.
Neglect quark masses and introduce energy fractions 
$x_j = 2E_j/E_{\mathrm{cm}}$ in the rest frame of the process.
Then the cross section is of the form
\begin{equation}
\frac{\mathrm{d} \sigma_{\mathrm{ME}}}{\sigma_0} = 
\frac{\alpha_{\mathrm{s}}}{2\pi} \, 
\frac{4}{3} \, \frac{x_1^2 + x_2^2}{(1-x_1)(1-x_2)} \, 
\mathrm{d} x_1 \, \mathrm{d} x_2 ~,
\end{equation}
where $\sigma_0$ is the $q \overline{q}$ cross section, i.e.\
without the gluon emission.

Now study the kinematics in the limit $x_2 \to 1$. Since
$1 - x_2 = m_{13}^2 / E_{\mathrm{cm}}^2$ we see that this 
corresponds to the ``collinear region'', where the separation 
between the $q$ and $g$ vanishes. Equivalently, the virtuality
$Q^2 = Q_i^2 = m_{13}^2$ of the intermediate quark propagator $i$ in 
Fig.~\ref{fig:qqg}a vanishes. Although the full answer contains
contributions from both graphs it is obvious that, in this region,
the amplitude of the one in Fig.~\ref{fig:qqg}a dominates over the 
one in Fig.~\ref{fig:qqg}b. We can therefore view the process
as $\gamma^*/Z^0 \to q \overline{q}$ followed by $q \to q g$. 
Define the energy sharing in the latter branching by 
$E_q = z E_i$ and $E_g = (1-z) E_i$. The kinematics relations then are
\begin{eqnarray}
& & 1 - x_2 = \frac{m_{13}^2}{E_{\mathrm{cm}}^2} = 
\frac{Q^2}{E_{\mathrm{cm}}^2} ~~ \Longrightarrow ~~
\mathrm{d} x_2 = \frac{\mathrm{d}Q^2}{E_{\mathrm{cm}}^2}\\
& & x_1 \approx z~~ \Longrightarrow ~~\mathrm{d} x_1 \approx \mathrm{d} z\\ 
& & x_3 \approx 1-z
\end{eqnarray}
so that
\begin{equation}
\mathrm{d}\mathcal{P} = \frac{\mathrm{d}\sigma_{\mathrm{ME}}}{\sigma_0} =
\frac{\alpha_{\mathrm{s}}}{2\pi} \; \frac{\mathrm{d} x_2}{(1-x_2)} \;
\frac{4}{3} \, \frac{x_2^2 + x_1^2}{(1-x_1)} \; \mathrm{d} x_1 
\approx \frac{\alpha_{\mathrm{s}}}{2\pi} \; \frac{\mathrm{d} Q^2}{Q^2} \; 
\frac{4}{3} \, \frac{1 + z^2}{1-z} \; \mathrm{d} z
\label{eq:Pqtoqg}
\end{equation}
Here $\mathrm{d} Q^2/Q^2$ corresponds to the ``collinear'' or ``mass''
singularity and $\mathrm{d} z / (1-z) = \mathrm{d}E_g / E_g$ to the
soft-gluon singularity.

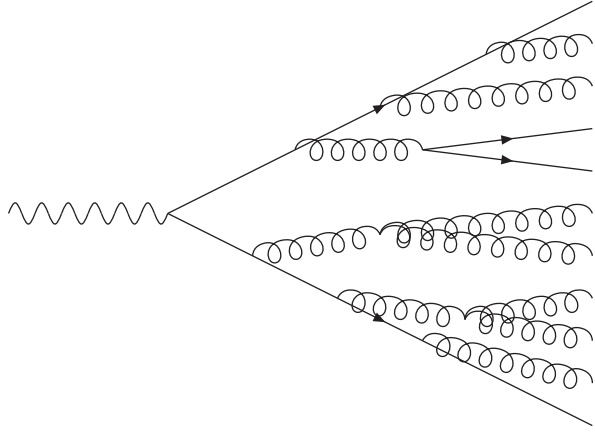
\begin{figure}[t]  
\begin{center}    
\begin{picture}(240,160)(-70,-80)
\Photon(-60,0)(0,0){4}{6}
\ArrowLine(0,0)(160,80)
\ArrowLine(0,0)(160,-80)
\Gluon(32,-16)(80,-8){4}{5}
\Gluon(80,-8)(160,-16){4}{8}\Gluon(80,-8)(160,0){4}{8}
\Gluon(64,-32)(112,-40){4}{5}
\Gluon(112,-40)(160,-32){4}{5}\Gluon(112,-40)(160,-48){4}{5}
\Gluon(96,-48)(160,-64){4}{7}
\Gluon(48,24)(96,24){4}{5}
\ArrowLine(96,24)(160,16)\ArrowLine(96,24)(160,32)
\Gluon(80,40)(160,48){4}{8}
\Gluon(120,60)(160,64){4}{4}
\end{picture}
\end{center}
\caption{A cascade of successive branchings.
\label{fig:fsrshower}}
\end{figure}

The interesting aspect of eq.~(\ref{eq:Pqtoqg}) is that it is universal:
whenever there is a massless quark in the final state, this equation
provides the probability for the same final state except for the quark 
being replaced by an almost collinear $q g$ pair (plus some other slight 
kinematics adjustments to conserve overall energy and momentum). That is 
reasonable: in a general process any number of distinct Feynman graphs
may contribute and interfere in a nontrivial manner, but once we go to
a collinear region only one specific graph will contribute, and that 
graphs always has the same structure, in this case with an intermediste
quark propagator. Corresponding rules can be derived for what happens
when a gluon is replaced by a collinear $gg$ or $q\overline{q}$ pair.
These rules are summarized by the DGLAP equations \cite{dglap}
\begin{eqnarray}
\mathrm{d}\mathcal{P}_{a\to bc} & = & \frac{\alpha_{\mathrm{s}}}{2\pi} 
\, \frac{\mathrm{d} Q^2}{Q^2} \, P_{a \to bc}(z) \, \mathrm{d} z
\label{eq:dglap} \\
\mathrm{where} 
& & P_{q \to qg}  =  \frac{4}{3} \, \frac{1 + z^2}{1-z} ~,\\
& & P_{g \to gg}  =  3 \, \frac{(1 - z(1-z))^2}{z(1-z)} ~,\\
& & P_{g \to q\overline{q}}  =   \frac{n_f}{2} \, (z^2 + (1-z)^2)~~~
(n_f = \mathrm{no.~of~quark~flavours}) ~.
\end{eqnarray}
Furthermore, the rules can be combined to allow for the successive 
emission in several steps, e.g.\ where a $q \to q g$ branching is
followed by further branchings of the daughters. That way a whole 
shower develops, Fig.~\ref{fig:fsrshower}.

Such a picture should be reliable in cases where the emissions are 
strongly ordered, i.e.\ $Q_1^2 \gg Q_2^2 \gg Q_3^2 \ldots$. Showers would 
not be useful if they only could be applied to strongly-ordered 
parton configurations, however. A further study of the 
$\gamma^*/Z^0 \to q \overline{q} g$ example shows that the simple sum 
of the $q \to q g$ and $\overline{q} \to \overline{q} g$ branchings 
reproduce the full matrix elements, with interference included, to better 
than a factor of 2 over the full phase space. This is one of the simpler 
cases, and of course one should expect the accuracy to be worse for
more complicated final states. Nevertheless, it is meaningful to 
use the shower over the whole strictly-ordered, but not necessarily 
strongly-ordered, region $Q_1^2 > Q_2^2 > Q_3^2 \ldots$ to obtain an 
approximate answer for multiparton topologies for which the complete
matrix elements would be too lengthy.

With the parton-shower approach, the big probability for one branching
$q \to qg$ turns into a big probability for several successive branchings.
Nevertheless we did not tame the fact that probabilities blow up in the
soft and collinear regions. For sure, perturbation theory will cease to 
be meaningful at so small $Q^2$ scales that $\alpha_{\mathrm{s}}(Q^2)$ 
diverges; there confimenent effects and hadronization phenomena take over. 
Typically therefore some lower cutoff at around 1~GeV is used to regulate
both soft and collinear divergences: below such a scale no further 
branchings are simulated. Whatever perturbative effects may remain 
are effectively pushed into the parameters of the nonperturbative 
framework. That way we avoid the singularities, but we can still have
``probabilities'' well above unity, which does not seem to make sense.  

This brings us to the second big concept of this section,
the \textit{Sudakov (form) factor} \cite{sudakov}.
In the context of particle physics it has a specific meaning related to 
the properties of the loop diagrams, but more generally we can just see 
it as a consequence of the conservation of total probability
\begin{equation}
\mathcal{P}(\mathrm{nothing~happens}) = 
1 - \mathcal{P}(\mathrm{something~happens}) ~,
\end{equation}
where the former is multiplicative in a time-evolution sense:
\begin{equation}
\mathcal{P}_{\mathrm{nothing}}(0 < t \leq T) =
\mathcal{P}_{\mathrm{nothing}}(0 < t \leq T_1) \;
\mathcal{P}_{\mathrm{nothing}}(T_1 < t \leq T) ~. 
\end{equation}
Now subdivide further, with $T_i = (i/n) T$, $0 \leq i \leq n$:
\begin{eqnarray}
\mathcal{P}_{\mathrm{nothing}}(0 < t \leq T) 
& = & \lim_{n \to \infty} \prod_{i = 0}^{n-1}
\mathcal{P}_{\mathrm{nothing}}(T_i < t \leq T_{i+1}) \nonumber \\
& = & \lim_{n \to \infty} \prod_{i = 0}^{n-1}
\left( 1 - \mathcal{P}_{\mathrm{something}}(T_i < t \leq T_{i+1}) \right)
 \nonumber \\
& = & \exp \left( - \lim_{n \to \infty} \sum_{i = 0}^{n-1}
\mathcal{P}_{\mathrm{something}}(T_i < t \leq T_{i+1}) \right)
 \nonumber \\
& = & \exp \left( - \int_0^T 
\frac{\mathrm{d} \mathcal{P}_{\mathrm{something}}(t)}{\mathrm{d} t} 
\mathrm{d} t \right) \nonumber \\[3mm]
\Longrightarrow ~~ \mathrm{d}\mathcal{P}_{\mathrm{first}}(T) 
& = & \mathrm{d}\mathcal{P}_{\mathrm{something}}(T) \; 
\exp \left( - \int_0^T \frac{\mathrm{d} \mathcal{P}_{\mathrm{something}}(t)}%
{\mathrm{d} t} \mathrm{d} t \right)  ~.
\end{eqnarray}
That is, the probability for something to happen for the \textit{first} 
time at time $T$ is the naive probability for this to happen, \textit{times} 
the probability that this did not yet happen. As such it applies to a host 
of situations. Take the example of football (relevant at the time of the
school). Assume that players are equally energetic and skillful from the
first minute of the match to the last. Then the chances of scoring a goal
is uniform in time, but the probability of scoring the \textit{first} goal
of the match is bigger at the beginning, because later on any goal could 
well be the second or third.
  
In physics a common example is that of radioactive decay. If the number
of undecayed radioactive nuclei at time $t$ is $\mathcal{N}(t)$, with
initial number $\mathcal{N}_0$ at time $t = 0$, then a naive ansatz
would be $\mathrm{d}\mathcal{N}/\mathrm{d}t = -c \mathcal{N}_0$, where
$c$ parametrizes the decay likelihood per unit of time. This equation has 
the solution $\mathcal{N}(t) = \mathcal{N}_0 (1 - ct)$, which becomes 
negative for $t > 1/c$, because by then the probability for having had a 
decay exceeds unity. So what we made wrong was not to take into account 
that only an undecayed nucleus can decay, i.e.\ that the equation ought to
have been $\mathrm{d}\mathcal{N}/\mathrm{d}t = -c \mathcal{N}(t)$
with the solution $\mathcal{N}(t) = \mathcal{N}_0 \exp(-ct)$. This is a 
nicely well-behaved expression, where the total probability for decays 
goes to unity only for $t \to \infty$. If $c$ had not been a constant but 
varied in time, $c = c(t)$, it is simple to show that the solution instead 
would have become 
\begin{equation}
\mathcal{N}(t) = \mathcal{N}_0 \exp \left( - \int_0^t c(t') \, 
\mathrm{d} t \right) ~~\Longrightarrow~~
\frac{\mathrm{d}\mathcal{N}}{\mathrm{d} t} = -c(t) \mathcal{N}_0    
\exp \left( - \int_0^t c(t') \, \mathrm{d} t \right) ~.
\end{equation}

For a shower the relevant ``time'' scale is something like $1/Q$, by the 
Heisenberg uncertainty principle. That is, instead of evolving to later 
and later times we evolve to smaller and smaller $Q^2$. Thereby the DGLAP 
eq.~(\ref{eq:dglap}) becomes
\begin{equation}
\mathrm{d}\mathcal{P}_{a\to bc} = \frac{\alpha_{\mathrm{s}}}{2\pi} \, 
\frac{\mathrm{d} Q^2}{Q^2} \, P_{a \to bc}(z) \, \mathrm{d} z \;
\exp \left( - \sum_{b,c} \int_{Q^2}^{Q_{\mathrm{max}}^2}  
\frac{\mathrm{d} {Q'}^2}{{Q'}^2} \int  \frac{\alpha_{\mathrm{s}}}{2\pi} \, 
P_{a \to bc}(z') \, \mathrm{d} z' \right) ~,
\end{equation}
where the exponent (or simple variants thereof) is the Sudakov factor.  
As for the radioactive-decay example above, the inclusion of a Sudakov
ensures that the total probability for a parton to branch never exceeds
unity. Then you may have sequential radioactive decay chains, and you may
have sequential parton branchings, but that is another story. 

It is a bit deeper than that, however. Just as the standard branching
expressions can be viewed as approximations to the complete matrix 
elements for real emission, the Sudakov is an approximation to the 
complete virtual corrections from loop graphs. The divergences in real 
and virtual emissions, so strange-looking in the matrix-element language, 
here naturally combine to provide a physical answer everywhere. What is 
not described in the shower, of course, is the non-universal finite parts 
of the real and virtual matrix elements.    

The implementation of a cascade evolution now makes sense. Starting 
from a simple $q\overline{q}$ system the $q$ and $\overline{q}$ are 
individually evolved downwards from some initial $Q_{\mathrm{max}}^2$ 
until they branch. At a branching the mother parton disappears and
is replaced by two daughter partons, which in their turn are evolved 
downwards in $Q^2$ and may branch. Thereby the number of partons
increases, until the lower cutoff scale is reached.

This does not mean that everything is uniquely specified. In particular, 
the choice of evolving in $Q^2 = m^2$ is by no means obvious. Any 
alternative variable $P^2 = f(z) \, Q^2$ would work equally well,
since $\mathrm{d} P^2 / P^2 = \mathrm{d} Q^2 / Q^2$. Alternative 
evolution variables therefore include the transverse momentum,
$p_{\perp}^2 \approx z(1-z)m^2$, and the energy-weighted emission angle
$E^2 \theta^2 \approx m^2 / (z (1-z))$.  

Both these two alternative choices are favourable when the issue 
of \textit{coherence} is introduced. Coherence means that emissions 
do not occur independently. For instance, consider 
$g_1 \to g_2 \, g_3$, followed by an emission of a gluon either 
from 2 or 3. When this gluon is soft it cannot resolve the 
individual colour charges of $g_2$ and $g_3$, but only the net charge
of the two, which of course is the charge of $g_1$. Thereby the 
multiplication of partons in a shower is reduced relative to naive
expectations. As it turns out, evolution in $p_{\perp}$ or angle 
automatically includes this reduction, while one in mass does not.

In the study of FSR, e.g.\ at LEP, three algorithms have been commonly
used. The \textsc{Herwig} angular-ordered and \textsc{Pythia} mass-ordered
ones are conventional parton showers as described above, while the
\textsc{Ariadne} \cite{ariadne} $p_{\perp}$-ordered one is 
based on a picture of dipole emissions. That is, instead of considering 
$a \to b \, c$ one studies $a \, b \to c\, d\, e$. One aspect of this is 
that, in addition to the branching parton, \textsc{Ariadne} also 
explicitly includes a ``recoil parton'' needed for overall 
energy--momentum conservation. Additionally emissions off $a$ and $b$
are combined in a well-defined manner.

All three approaches have advantages and disadvantages. As already
mentioned, \textsc{Pythia} does not inherently include coherence, 
but has to add that approximately by brute force. Both \textsc{Pythia}
and \textsc{Herwig} break Lorentz invariance slightly. 
The \textsc{Herwig} algorithm cannot cover the full phase space with 
it emissions, but has to fill in some ``dead zones'' using higher-order 
matrix elements. The \textsc{Ariadne} dipole picture does not include 
$g \to q \overline{q}$ branchings in a natural way. And so on. 

When all is said and done, it turns out that all three algorithms do a
quite decent job of describing LEP data \cite{lepdata}, but typically 
\textsc{Ariadne} does best and \textsc{Herwig} worst. Since 
\textsc{Ariadne} uses \textsc{Pythia} for hadronization the difference 
between those two is entirely due to the shower algorithms, while
comparisons with \textsc{Herwig} also are complicated by significant 
differences in hadronization. 

\subsection{Initial-state radiation}

The structure of initial-state radiation (ISR) is more complicated
than that of FSR, since the nontrivial structure of the incoming
hadrons enter the game. A proton 
is made up out of three quarks, $u u d$, plus the gluons that bind
them together. This picture is not static, however: gluons are 
continuously emitted and absorbed by the quarks, and each gluon may 
in its turn temporarily split into two gluons or into a $q \overline{q}$ 
pair. Thus a proton is teeming with activity, and much of it in a 
nonperturbative region where we cannot calculate. We are therefore
forced to introduce the concept of a parton density $f_b(x, Q^2)$
as an empirical distribution, describing the probability to find a
parton of species $b$ in a hadron, with a fraction $x$ of the hadron 
energy--momentum when the hadron is probed at a resolution scale $Q^2$.

While $f_b(x, Q^2)$ itself cannot be predicted, the change of $f_b$ 
with resolution scale can, once $Q^2$ is large enough that perturbation
theory should be applicable:
\begin{equation}
\frac{\mathrm{d} f_b(x,Q^2)}{\mathrm{d}(\ln Q^2)} = \sum_a
\int_x^1 \frac{\mathrm{d} z}{z} \, f_a(x',Q^2) \, 
\frac{\alpha_{\mathrm{s}}}{2\pi} \, 
P_{a \to bc} \left(z = \frac{x}{x'} \right) ~.
\label{eq:dglapisr} 
\end{equation}
This is actually nothing but our familiar DGLAP equations. Before they
were written in an exclusive manner: given a parton $a$, what is the
probability that it will branch to $b \, c$ during a change 
$\mathrm{d}Q^2$? Here the formulation is instead inclusive: given 
that the probability distributions $f_a(x, Q^2)$ of all partons $a$ 
are known at a scale $Q^2$, how is the distribution of partons $b$
changed by the set of possible branchings $a \to b$ ($+c$, here implicit). 
The splitting kernels $P_{a \to bc}(z)$ are the same to leading order, 
but differ between ISR and FSR in higher orders. In higher orders also 
the concept of $f_b(x,Q^2)$ as a positive definite probability is lost, 
additional complications that we will not consider any further here.

Even though eqs.~(\ref{eq:dglap}) and (\ref{eq:dglapisr}) are equivalent, 
the physics context is different. In FSR the outgoing partons have been 
kicked to large timelike virtualities by the hard process and then cascade 
downwards towards the mass shell. In ISR we rather start out with a simple 
proton at early times and then allow more and more spacelike virtualities 
as we get closer to the hard interaction. Not that big fluctuations could 
not happen at early times --- they do --- but if they happen too early
the uncertainty relation does not allow them to live long enough to
be of any intererest to us. That is, the higher the virtuality, the
later the fluctuation has to occur. 

\begin{figure}[t]  
\begin{center}    
\begin{picture}(420,120)(-10,-60)
\SetWidth{2}
\ArrowLine(0,0)(80,0)
\ArrowLine(80,0)(160,-10)
\ArrowLine(160,-10)(240,10)
\ArrowLine(240,10)(320,-20)
\ArrowLine(320,-20)(400,20)
\Line(395,15)(405,25)
\Line(395,25)(405,15)
\SetWidth{1}
\ArrowLine(80,0)(120,10)
\DashArrowLine(120,10)(160,0){5}
\DashArrowLine(120,10)(160,30){5}
\DashArrowLine(160,30)(200,20){5}
\DashArrowLine(160,30)(200,40){5}
\ArrowLine(160,-10)(200,-30)
\DashArrowLine(200,-30)(240,-20){5}
\DashArrowLine(240,-20)(280,-15){5}
\DashArrowLine(240,-20)(280,-30){5}
\DashArrowLine(200,-30)(240,-40){5}
\DashArrowLine(240,-40)(280,-45){5}
\DashArrowLine(240,-40)(280,-60){5}
\ArrowLine(240,10)(280,40)
\DashArrowLine(280,40)(320,50){5}
\DashArrowLine(280,40)(320,20){5}
\DashArrowLine(320,20)(360,10){5}
\DashArrowLine(320,20)(360,40){5}
\DashArrowLine(360,40)(400,30){5}
\DashArrowLine(360,40)(400,50){5}
\ArrowLine(320,-20)(350,-40)
\DashArrowLine(350,-40)(400,-60){5}
\DashArrowLine(350,-40)(380,-20){5}
\DashArrowLine(380,-20)(400,-35){5}
\DashArrowLine(380,-20)(400,-5){5}
\end{picture}
\end{center}
\caption{A cascade of successive branchings. The thick line represents
the main chain of spacelike partons leading in to the hard interaction
(marked by a cross). The thin lines are partons that cannot be recombined,
while dashed lines are further fluctuations that may (if spacelike)
or may not (if timelike) recombine. In this graph lines can represent 
both quarks and gluons.
\label{fig:isrshower}}
\end{figure}
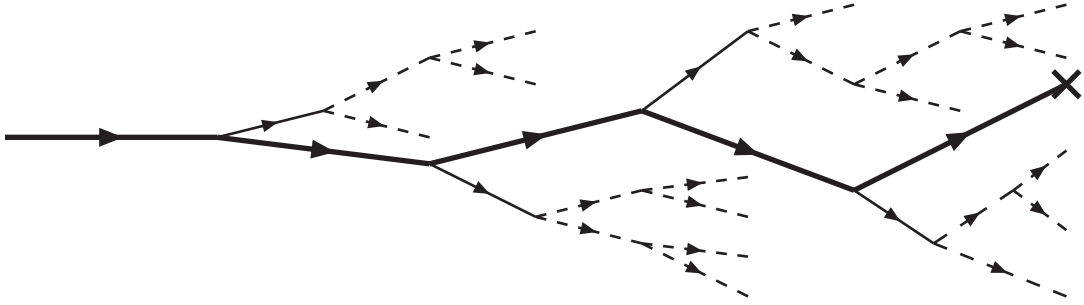

So, when the hard scattering occurs, in some sense the initial-state
cascade is already there, as a virtual fluctuation. Had no collision
occured the fluctuation would have collapsed back, but now one of the
partons of the fluctuation is kicked out in a quite different direction
and can no longer recombine with its sister parton from its last
branching, nor with its aunt from the last-but-one branching. And so on 
for each preceding branching in the cascade that lead up to this 
particular parton. Post facto we therefore see that a chain of 
branchings with increasing $Q^2$ values built up an ISR shower,
Fig.~\ref{fig:isrshower}.
  
The obvious way to simulate this situation would be to pick partons
in the two incoming hadrons from parton densities at some low $Q^2$ scale, 
and then use the exclusive formulation of eq.~(\ref{eq:dglap}) to construct 
a complete picture of partons available at higher $Q^2$ scales, event
by event. The two sets of incoming partons could then be weighted by
the cross section for the process under study. A problem is that this
may not be very efficient. We have to evolve for all possible fluctuations, 
but at best one particular parton will collide and most of the other
fluctuations will collapse back. The cost may become prohibitive when 
the process of interest has a constrained phase space, like a light-mass 
Higgs which has to have the colliding partons matched up in a very
narrow mass bin. 

There are ways to speed up this ``forwards evolution'' approach.
However, the most common solution is instead to adopt a 
``backwards evolution'' point of view \cite{backwards}. Here one starts 
at the hard interaction and then tries to reconstruct what 
happened ``before''. To be more precise, the cross-section formula in 
eq.~(\ref{eq:sigma}) already includes the summation over all possible
incoming shower histories by the usage of $Q^2$-dependent parton
densities. Therefore what remains is to pick one exclusive shower
history from the inclusive set that went into the $Q^2$-evolution.
To do this, recast eq.~(\ref{eq:dglapisr}) as 
\begin{equation} 
\mathrm{d}\mathcal{P}_b = \frac{\mathrm{d} f_b}{f_b} = 
|\mathrm{d}(\ln Q^2) | \, \sum_a \int \mathrm{d} z \, 
\frac{x' f_a(x',t)}{x f_b(x,t)} \, \frac{\alpha_{\mathrm{s}}}{2\pi} 
\, P_{a \to bc} \left(z = \frac{x}{x'} \right) ~.
\end{equation}
Then we have defined a \textit{conditional probability}: if parton 
$b$ is present at scale $Q^2$, what is the probability that it will 
turn out to have come from a branching $a \to b \, c$ at some 
infinitesimally \textit{smaller} scale? (Recall that the original
eq.~(\ref{eq:dglapisr}) was defined for increasing virtuality.)
Like for FSR this expression has to be modified by a Sudakov factor 
to preserve total probability, and this factor is again the exponent 
of the real-emission expression with a negative sign, integrated 
over $Q^2$ from an upper starting scale $Q_{\mathrm{max}}^2$ down to 
the $Q^2$ of the hypothetical branching.

The approach is now clear. First a hard scattering is selected, making
use of the $Q^2$-evolved parton densities. Then, with the hard process
as upper maximum scale, a succession of ISR branchings are reconstructed
at lower and lower $Q^2$ scales, going ``backwards in time'' towards
the early low-virtuality initiators of the cascades. Again some 
cutoff needs to be introduced when the nonperturbative regime 
is reached. 
 
Unfortunately the story does not end there. For FSR we discussed the 
need to take into account coherence effects and the possibility to use 
different variables. Such issues exist here as well, but also additional
ones. For instance, evolution need not be strictly ordered in $Q^2$.
(Eq.~(\ref{eq:ptkin}) only gives you $Q_b^2 > z Q_a^2$, not 
$Q_b^2 > Q_a^2$.) Non-ordered chains in some cases can be important. 
Another issue is that there can be so many partons evolving inside a 
hadron that they become close-packed, which leads to additional 
recombinations. See the lectures of G.~Ingelman for further details 
\cite{ingelman}.

In summary, ISR and FSR share many aspects, but also differ. The DGLAP
evolution with Sudakov factors allows a simple probabilistic framework,
where an initial parton undergoes successive branchings. For ISR the 
branching (usually) is in terms of higher and higher spacelike virtualities 
as the hard scattering approaches, while for FSR the branchings
involve lower and lower timelike virtualities as the hard scattering 
recedes. In FSR both daughter partons appear on equal footing, 
in that both can be timelike and branch further. In ISR only the daughter
parton on its way in to the hard scattering can be spacelike virtual; 
its sister will become part of the final state and thus has to be on
mass shell, or else be timelike and start an FSR shower of its own.
And, last but not least, the whole FSR framework is considerably better
understood than the ISR one.
 
\section{Combining matrix elements and parton showers}

As we have seen, both matrix elements (ME) and parton showers (PS)
have advantages and disadvantages. 

To recall, ME allow a systematic 
expansion in powers of $\alpha_{\mathrm{s}}$, and thereby offer a 
controlled approach towards higher precision. Calculations can be done 
with several (up to $\sim 8$) partons in the final state, so long as only 
Born-level results are asked for, and it is possible to tailor the 
phase-space cuts for these partons precisely to the experimental needs.
Loop calculations are much more difficult, on the other hand, and the
mathematically correct cancellation between real- and virtual-emission
graphs in the soft/collinear regions is not physically sensible. 
Therefore ME cannot be used to explore the internal structure of a jet, 
and are difficult to match to hadronization models, which are supposed 
to take over in the very soft/collinear region. 

PS, on the other hand, clearly are approximate and do not come with a 
guaranteed level of precision for well separated jets. You cannot steer 
the probabilistic evolution of a shower too much, and therefore the 
efficiency for obtaining events in a specific region of phase space 
can be quite low. On the other hand, PS are universal, so for any
new model you only need to provide the basic hard process and then PS
will turn that into reasonably realistic multiparton topologies.
The use of Sudakov factors ensures a physically sensible behaviour
in the soft/collinear regions, and it is also here that the PS formalism
is supposed to be as most reliable. It is therefore possible to obtain 
a good picture of the internal structure of jets, and to provide a 
good match to hadronization models.

In a nutshell: ME are good for well separated jets, PS for the structure
inside jets. Clearly the two complement each other, and a marriage is 
highly desirable. To do this, without doublecounting or gaps in the
phase space coverage, is less trivial, and several alternative approaches
have been developed. In the following we will discuss three main options:
merging, vetoed parton showers and MC@NLO, roughly ordered in increasing
complexity. Which of these to use may well depend on the task at hand.

\subsection{Merging}

The aspiration of merging is to cover the whole phase space with a smooth
transition from ME to PS. The typical case would be a process where the
lowest-order (LO) ME is known, as well as the next-to-leading-order (NLO) 
real-emission one, say of an additional gluon. The shower should then 
reproduce
\begin{equation} 
W^{\mathrm{ME}} = \frac{1}{\sigma(\mathrm{LO})} ~~
\frac{\mathrm{d}\sigma(\mathrm{LO}+ g)}{\mathrm{d}(\mathrm{phasespace})}
\label{eq:wme}
\end{equation}
starting from a LO topology. If the shower populates phase space 
according to $W^{\mathrm{PS}}$ this implies that a correction factor
$W^{\mathrm{ME}} / W^{\mathrm{PS}}$ need to be applied. 

At first glance this does not apper to make sense: if all we do is get 
back $W^{\mathrm{ME}}$, then what did we gain? However, the trick is
to recall that the PS formula comes in two parts: the 
real-emission answer and a Sudakov factor that ensures total conservation
of probability. What we have called $W^{\mathrm{PS}}$ above 
should only be the real-emission part of the story. It is also this 
one that we know \textit{will} agree with $W^{\mathrm{ME}}$ in the
soft and collinear regions. Actually, with some moderate amount of effort 
it is often possible to ensure that $W^{\mathrm{ME}} / W^{\mathrm{PS}}$
is of order unity over the whole phase space, and to adjust the showers 
in the hard region so that the ratio always is below unity, i.e.\
so that standard Monte Carlo rejection techniques can be used. What the 
Sudakov factor then does is introduce some ordering variable $Q^2$, so that 
the whole phase space is covered starting from ``hard'' emissions and 
moving to ``softer'' ones. At the end of the day this leads to a 
distribution over phase space like
\begin{equation} 
W^{\mathrm{PS}}_{\mathrm{actual}}(Q^2) =
W^{\mathrm{ME}}(Q^2)  
\exp \left( - \int_{Q^2}^{Q_{\mathrm{max}}^2} 
W^{\mathrm{ME}}({Q'}^2) \, \mathrm{d}{Q'}^2 \right) ~. 
\end{equation}
That is, we have used the PS choice of evolution variable to provide an
exponentiated version of the ME answer. As such it agrees with the ME
answer in the hard region, where the Sudakov factor is close to unity,
and with the PS in the soft/collinear regions, where
$W^{\mathrm{ME}} \approx W^{\mathrm{PS}}$.  

The method is especially convenient for resonance decays, such as
$e^+ e^- \to \gamma^*/Z^0 \to q \overline{q}$ where it was first
introduced \cite{mats}. In that case there is an added bonus: the 
full NLO answer, with virtual corrections included, is known to be
$\sigma^{\mathrm{NLO}} = \sigma^{\mathrm{LO}} \, 
(1 + \alpha_{\mathrm{s}} / \pi)$. So it is trivial to use the procedure
above and rescale everything by $(1 + \alpha_{\mathrm{s}} / \pi)$ to
obtain a complete NLO answer. Note that the difference between using 
$\sigma(\mathrm{LO})$ or $\sigma(\mathrm{NLO})$ in the denominator
of eq.~(\ref{eq:wme}) only gives a difference to 
$\mathcal{O}(\alpha_{\mathrm{s}}^2)$, i.e. to NNLO.

In \textsc{Pythia} this approach is used for essentially all resonance
decays in the Standard Model and minimal supersymmetric extensions
thereof: $t \to b W^+$, $W^+ \to u \overline{d}$, $H \to b \overline{b}$,
$\chi^0 \to \tilde{q} \overline{q}$, $\tilde{q} \to q \tilde{g}$, \ldots
\cite{emanuel}. It is also used in ISR to describe 
$q\overline{q} \to \gamma^*/Z^0/W^{\pm}$ \cite{gabriela}, but here 
the NLO corrections are more tricky, so the cross section remains as 
provided by the LO number. 

Merging is also used for several processes in \textsc{Herwig}, such as
$\gamma^*/Z^0 \to q \overline{q}$, $t \to b W^+$ and 
$q\overline{q} \to \gamma^*/Z^0/W^{\pm}$ \cite{hwmatch}.
A special problem here is that the angular-ordered algorithms, both
for FSR and for ISR, leave some ``dead zones'' of hard emissions that 
are kinematically forbidden for the shower to populate. It is therefore 
necessary to start directly from higher-order matrix elements in these 
regions. A consistent treatment still allows a smooth joining across
the boundary.
 
\subsection{Vetoed parton showers}

In some sense vetoed parton showers is an extension of the merging
approach above. The objective is still to combine the real-emission
behaviour of ME with the emission-ordering-variable-dependent
Sudakov factors of PS. While the merging approach only works 
for combining the LO and NLO expressions, however, the vetoed parton
showers offer a generic approach for combining several different
orders. Therefore it is likely to be a standard tool for many studies
in the future. 

To understand how the algorithm works, consider a lowest-order process
such as $q \overline{q} \to W^{\pm}$. For each higher order one 
additional jet would be added to the final state, so long as only 
real-emission graphs are considered: in first order e.g.\
$q \overline{q} \to W^{\pm} g$, in second order e.g.\
$q \overline{q} \to W^{\pm} g g$, and so on. Call these (differential)
cross sections $\sigma_0$, $\sigma_1$, $\sigma_2$, \ldots. It should 
then come as no surprise that each $\sigma_i$, $i \geq 1$, contains
soft and collinear divergences. We therefore need to impose some
set of ME phase-space cuts, e.g. on invariant masses of parton pairs,
or on parton energies and angular separation between them. When these
cuts are varied, so that e.g. the mass or energy thresholds are lowered
towards zero, all of these $\sigma_i$, $i \geq 1$, increase without 
bounds.
 
However, in the ME approach without virtual corrections there is no
``detailed balance'', wherein the addition of cross section to 
$\sigma_{i+1}$ is compensated by a depletion of $\sigma_i$. That is, if 
you have an event with $i$ jets at some resolution scale, and a lowering
of the minimal jet energy reveals the presence of one additional
jet, then you should reclassify the event from being $i$-jet to being
$i+1$-jet. Add one, subtract one, with no net change in 
$\sum_i \sigma_i$. So the trick is to use the Sudakovs of showers 
to ensure this detailed balance. Of course, in a complete description
the cancellation between real and virtual corrections is not completely
exact but leaves a finite net contribution, which is not predicted
in this approach.

A few alternative algorithms exist along these lines.  
All share the three first steps as follows \cite{mlm}:\\
1) Pick a hard process within the ME-cuts-allowed phase-space region,
in proportions provided by the ME integrated over the respective
allowed region, $\sigma_0 : \sigma_1 : \sigma_2 : \ldots$.
Use for this purpose a fix $\alpha_{\mathrm{s}0}$ larger than the 
$\alpha_{\mathrm{s}}$ values that will be used below.\\
2) Reconstruct an imagined shower history that describes how the 
event could have evolved from the lowest-order process to the 
actual final state. That provides an ordering of emissions by
whatever shower-evolution variable is intended.\\
3) The ``best-bet'' choice of $\alpha_{\mathrm{s}}$ scale in
showers is known to be the squared transverse momentum of the 
respective branching. Therefore a factor 
$W_{\alpha} = \prod_{\mathrm{branchings}} 
(\alpha_{\mathrm{s}}(p_{\perp i}^2) / \alpha_{\mathrm{s}0})$,
provides the probability that the event should be retained.
 
Now the algorithms part way. In the CKKW--L approach the subsequent
steps are:\\
4) Evaluate Sudakov factors for all the ``propagator'' lines in the 
shower history reconstructed in step 2, i.e. for intermediate partons 
that split into further partons, and also for the evolution of the 
final partons down to the ME cuts without any further 
emissions. This provides an acceptance weight
$W_{\mathrm{Sud}} = \prod_{\mathrm{``propagators''}} 
\mathrm{Sudakov}(Q^2_{\mathrm{beg}}, Q^2_{\mathrm{end}})$
where $Q^2_{\mathrm{beg}}$ is the large scale where a parton is
produced by a branching and $Q^2_{\mathrm{end}}$ either is the 
scale at which the parton branches or the ME cuts, the 
case being.\\
4a) In the CKKW approach \cite{ckkw} the Sudakovs are evaluated by 
analytical formulae, which is fast.\\
4b) In the L approach \cite{leif} trial showers are used to evaluate
Sudakovs, which is slower but allows a more precise modelling
of kinematics and phase space than offered by the analytic expression.\\
5) Now the matrix-element configuration can be evolved further,
to provide additional jets below the ME cuts used. In order to avoid
doublecounting of emissions, any branchings that might occur above
the ME cuts must be vetoed.

The MLM approach \cite{mlm} is rather different. Here the steps instead 
are:\\
4') Allow a complete parton shower to develop from the selected parton 
configuration.\\
5') Cluster these partons back into a set of jets, e.g. using a cone-jet
algorithm, with the same jet-separation criteria as used when the
original parton configuration was picked.\\
6') Try to match each jet to its nearest original parton.\\
7') Accept the event only if the number of clustered jets agrees with 
the number of original partons, and if each original parton is sensibly 
matched to its jet. This would not be the case e.g.\ if one parton gave 
rise to two jets, or two partons to one jet, or an original $b$ quark 
migrated outside of the clustered jet.\\
The point of the MLM approach is that the probability of \textit{not} 
generating any additional fatal jet activity during the shower evolution 
is provided by the Sudakovs used in the step 4'. 

\subsection{MC@NLO}

MC@NLO \cite{mcnlo} in some respects is the most ambitious approach:
it aims to get not only real but also virtual contributions correctly
included, so that cross sections are accurate to NLO, and that NLO
results are obtained for all observables when formally expanded in
powers of $\alpha_{\mathrm{s}}$. Thus hard emissions should again
be generated according to ME, while soft and collinear ones should 
fall within the PS regime.

In simplified terms, the scheme works as follows:\\
1) Calculate the NLO ME corrections to an $n$-body process,
including $n+1$-body real corrections and $n$-body virtual ones.\\
2) Calculate analytically how a first branching in a shower
starting from a $n$-body topology would populate $n+1$-body phase 
space, excluding the Sudakov factor.\\
3) Subtract the shower expression from the $n+1$ ME one to obtain the
``true'' $n+1$ events, and consider the rest as belonging to the 
$n$-body class. The PS and ME expressions agree in the soft
and collinear limits, so the singularities there cancel, leaving
finite cross sections both for the $n$- and $n+1$-body event 
classes.\\
4) Now add showers to both kinds of events.

\begin{figure}[t]  
\begin{center}    
\begin{picture}(350,200)(-50,-50)
\LongArrow(0,0)(180,0)\Text(190,0)[]{$p_{\perp Z}$}
\LongArrow(0,0)(0,135)\Text(0,145)[]{$\mathrm{d}\sigma / 
\mathrm{d} p_{\perp Z}$}
\Text(190,140)[l]{\framebox{simplified  example}}
\Curve{(13.3,133)(16.7,107)(21.4,83)(26.7,67)(33.3,53.3)(41.7,42.7)(53.3,33.3)%
(67,26.3)(83,21.3)(107,16.7)(133,13)(167,10)}
\LongArrow(70,98)(40,45)
\Text(60,110)[l]{$Z + 1$ jet ME}
\LongArrow(165,35)(90,10)
\Text(175,35)[l]{generate as Z + 1 jet + shower}
\Curve{(13.3,130)(16.7,103)(21.4,77)(26.7,59)(33.3,41)(40,26.7)(47,15.7)%
(53,6.7)(60,0)}
\LongArrow(165,110)(43,26)
\Text(175,110)[l]{$Z + 1$ jet according to shower}
\Text(175,95)[l]{(first emission, without Sudakov)}
\LongArrow(165,60)(35,10)
\Text(175,60)[l]{generate as $Z$ + shower}
\SetWidth{2}
\LongArrow(0,0)(0,70)
\Text(-10,40)[r]{LO real}
\LongArrow(0,0)(0,-50)
\Text(-10,-30)[r]{NLO virtual}
\end{picture}
\end{center}
\caption{MC@NLO applied to a $Z^0$ production. The region between the 
two curves is considered as ``true'' $Z$ + 1 jet events, with  showers
added to that. The rest --- LO real, NLO virtual and NLO real in the
shower approximation --- are combined to cancel singularities and then
showered as simple $Z$ events.
\label{fig:mcnlo}}
\end{figure}
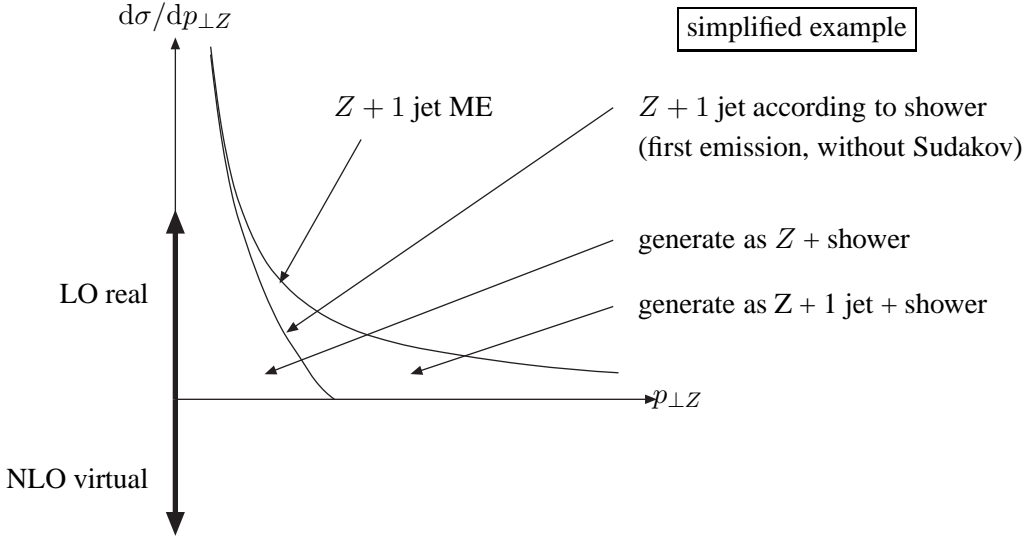

A toy example, for the case of $Z^0$ production, is shown in 
Fig.~\ref{fig:mcnlo}. Several more complicated processes have been 
considered, such as $b\overline{b}$, $t\overline{t}$ and $W^+W^-$
production. A technical problem is that, although ME and PS converge
in the collinear region, it is not guaranteed that ME is everywhere
above PS. This is solved by having a small fraction of events with
negative weights. 

In summary, MC@NLO is superiour in that it does provide the total
cross section for a process to NLO accuracy, and so is essential for
a set of precision tests. The real-emission $n+1$-body part is the same 
as used for merging, however, so for normalized event shapes the merging 
approach is as valid. (Any differences are of higher order.) Finally,
if multijet topologies need to be studied, where several orders may
contribute, vetoed showers are more appropriate. Each tool to its 
task.

\section{Multiple interactions}

The cross section for $2 \to 2$ QCD parton processes is dominated by 
$t$-channel gluon exchange, as we already mentioned, and thus diverges
like $\mathrm{d}p_{\perp}^2 / p_{\perp}^4$ for $p_{\perp} \to 0$.
Introduce a lower cut $p_{\perp \mathrm{min}}$ and integrate the
interaction cross section above this, properly convoluted with parton 
densities. At LHC energies this 
$\sigma_{\mathrm{int}}(p_{\perp \mathrm{min}})$ reaches
around 100~mb for $p_{\perp \mathrm{min}} = 5$~GeV, and 1000~mb at
around 2~GeV. Since each interaction gives two jets to lowest
order, the jet cross section is twice as big. This should be compared 
with an expected \textit{total} cross section of the order of 100~mb.
(QCD is a confining theory and thus provides a finite total cross
section, unlike QED, where infinitely small scattering angles are 
allowed at infinitely large distances.) In addition, at least a 
third of the total cross section is related to elastic scattering 
$p p \to p p$ and low-mass diffractive states $p p \to p X$ that could
not contain jets.

So can it really make sense that 
$\sigma_{\mathrm{int}}(p_{\perp \mathrm{min}}) > \sigma_{\mathrm{tot}}$? 
Yes, it can! The point is that each incoming hadron is a bunch of partons.
You can have several (more or less) independent parton--parton interactions
when these two bunches pass through each other. And the point is
that an event with $n$ interactions above $p_{\perp \mathrm{min}}$
counts once for the total cross section but once 
\textit{for each interaction} when the interaction rate is calculated.
That is,
\begin{equation}
\sigma_{\mathrm{tot}} = \sum_{n = 0}^{\infty} \sigma_n 
~~~~~~~~~~\mathrm{while}~~~~~~~~~~ 
\sigma_{\mathrm{int}} = \sum_{n = 0}^{\infty} n \, \sigma_n ~,
\end{equation} 
where $\sigma_n$ is the cross section for events with $n$ interactions.
Thus $\sigma_{\mathrm{int}} > \sigma_{\mathrm{tot}}$ is equivalent to
$\langle n \rangle > 1$, that each event on the average contains more 
than one interaction. Furthermore, if interactions really do occur
independently when two hadron pass by each other, then one would 
expect a Poissonian distribution,
$\mathcal{P}_n = \langle n \rangle^n \, \exp(-\langle n \rangle) / n!$,
so that several interactions could occur occasionally also when 
$\sigma_{\mathrm{int}}(p_{\perp \mathrm{min}}) < \sigma_{\mathrm{tot}}$,
e.g. for a larger $p_{\perp \mathrm{min}}$ cut. Energy--momentum
conservation ensures that interactions never are truly independent, 
and also other effects enter (see below), but the Poissonian ansatz
is still a useful starting point.

Multiple interactions (MI) can only be half the solution, however. The 
divergence for $p_{\perp \mathrm{min}} \to 0$ would seem to imply an 
infinite average number of interactions. But what one should 
realize is that, in order to calculate the 
$\mathrm{d}\hat{\sigma} / \mathrm{d}\hat{t}$ matrix elements
within standard perturbation theory, it has to be assumed that
free quark and gluon states exist at negative and positive infinity.
That is, the confinement of colour into hadrons of finite size
has not been taken into account. So obviously perturbation theory
has to have broken down by 
\begin{equation} 
p_{\perp \mathrm{min}} \simeq \frac{\hbar}{r_p} \approx 
\frac{0.2~\mathrm{GeV}\cdot\mathrm{fm}}{0.7~\mathrm{fm}} 
\approx 0.3~\mathrm{GeV} \simeq \Lambda_{\mathrm{QCD}} ~.
\end{equation}
The nature of the breakdown is also easy to understand: a 
small-$p_{\perp}$ gluon, to be exchanged between the two incoming hadrons, 
has a large transverse wavelength and thus almost the same phase across 
the extent of each hadron. The contributions from all the colour charges
in a hadron thus add coherently, and that means that they add to zero
since the hadron is a colour singlet. 

What is then the typical scale of such colour screening effects, 
i.e.\ at what $p_{\perp}$ has the interaction rate dropped to $\sim$half
of what it would have been if the quarks and gluons of a proton
had all been free to interact fully independently? That ought to be
related to the typical separation distance between a given colour 
and its opposite anticolour. When a proton contains many partons
this characteristic screening distance can well be much smaller than
the proton radius. Empirically we need to introduce a 
$p_{\perp \mathrm{min}}$ scale of the order of 2~GeV to describe 
Tevatron data, i.e. of the order of 0.1~fm separation. It is not
meaningful to take this number too seriously without a detailed model 
of the space--time structure of a hadron, however.

The 2~GeV number is very indirect and does not really tell exactly how
the dampening occurs. One can use a simple recipe, with a step-function
cut at this scale, or a physically more reasonable dampening by a factor
$p_{\perp}^4 / (p_{\perp 0}^2 + p_{\perp})^2$, plus a corresponding shift 
of the $\alpha_{\mathrm{s}}$ argument,
\begin{equation}
\frac{\mathrm{d} \hat{\sigma}}{\mathrm{d} p_{\perp}^2} 
\propto \frac{\alpha_{\mathrm{s}}^2(p_{\perp}^2)}{p_{\perp}^4}
\to  \frac{\alpha_{\mathrm{s}}^2(p^2_{\perp 0} + p_{\perp}^2)}%
{(p^2_{\perp 0} + p_{\perp}^2)^2} ~,
\label{eq:smoothmi}
\end{equation}
with $p_{\perp 0}$ a dampening scale that also lands at around 2~GeV.
This translates into a typical number of 2--3 interactions per event
at the Tevatron and 4--5 at LHC. For events with jets or other hard 
processes the average number is likely to be higher.

\subsection{Multiple-interactions models}

The first studies of complete events based on perturbatively generated 
MI \cite{maria} started out from a mimimally simple model:\\
1) Use a sharp $p_{\perp \mathrm{min}}$ cut as the key tunable parameter.\\
2) Only address inelastic nondiffractive events, i.e.\ with 
$\sigma_{\mathrm{nd}} \simeq (1/2 - 2/3) \sigma_{\mathrm{tot}}$, 
so that the average number of interactions per such events is
$\langle n \rangle = \sigma_{\mathrm{int}}(p_{\perp \mathrm{min}}) /
\sigma_{\mathrm{nd}}$.\\
3) To first approximation this gives a Poissonian distribution in the
number of interactions per event, with a fraction 
$\mathcal{P}_0 = e^{-\langle n \rangle}$ of purely low-$p_{\perp}$
interactions.\\
4) The interactions are generated in an ordered sequence of decreasing
$p_{\perp}$ values: $p_{\perp 1} > p_{\perp 2} > p_{\perp 3} > \ldots$.
This is possible with the standard Sudakov kind of trick:
\begin{equation}
\frac{\mathrm{d}\mathcal{P}}{\mathrm{d} p_{\perp i}} 
= \frac{1}{\sigma_{\mathrm{nd}}} 
\frac{\mathrm{d}\sigma}{\mathrm{d} p_{\perp}} \;
\exp\left[ - \int_{p_{\perp}}^{p_{\perp(i-1)}} 
\frac{1}{\sigma_{\mathrm{nd}}} 
\frac{\mathrm{d}\sigma}{\mathrm{d}p_{\perp}'} 
\mathrm{d}p_{\perp}' \right] ~,
\end{equation} 
with a starting $p_{\perp 0} = E_{\mathrm{cm}}/2$.\\
5) The ordering of emissions allows parton densities to be rescaled 
in $x$ after each interaction, so that energy--momentum is not violated.
Thereby the actual distribution of the number of interactions becomes
narrower than Poissonian, since the rate of a further interaction is 
reduced if previous ones already took away energy.\\ 
6) For technical reasons the model was simplified after the first
interaction, so that there only $gg$ or $q\overline{q}$ outgoing pairs
were allowed, and no showers were added to these additional $2 \to 2$
interactions.\\
Already this simple \textsc{Pythia}-based model is able to ``explain'' 
a large set of experimental data. 

Some additional features were then included in a more sophisticated 
variant.\\
1) Use the smooth turnoff of eq.~(\ref{eq:smoothmi}) all the way down to 
$p_{\perp} = 0$. Then an event has to contain at least one interaction
to be an event at all, i.e.\ $p_{\perp 0}$ has to be selected sufficiently
small that $\sigma_{\mathrm{int}} > \sigma_{\mathrm{nd}}$.\\  
2) Hadrons are extended, and therefore partons are distributed in 
(transverse) coordinates. To allow a flexible parametrization 
and yet have an easy-to-work-with expression, a double Gaussian 
$\rho_{\mathrm{matter}}(\mathbf{r}) 
= N_1 \exp \left( - r^2 / r_1^2 \right) +
N_2 \exp \left( - r^2 / r_2^2 \right)$ is used, where $N_2/N_1$ and
$r_2/r_1$ are tunable parameters.\\
3) The matter overlap during a collision, calculated by
\begin{equation} 
\mathcal{O}(b) = \int \mathrm{d}^3\mathbf{x} \, \mathrm{d} t \; 
\rho_{1,\mathrm{matter}}^{\mathrm{boosted}}(\mathbf{x}, t)
\rho_{2,\mathrm{matter}}^{\mathrm{boosted}}(\mathbf{x}, t) ~,
\end{equation}
directly determines the average activity in events at different 
impact parameter $b$: $\langle n(b) \rangle \propto \mathcal{O}(b)$.
That is, central collisions tend to have more activity, peripheral less,
but of course properly normalized so that the $b$-integrated interaction
cross section agrees with standard perturbation theory (modulo the 
already-discussed dampening at small $p_{\perp}$). Thereby the 
$\mathcal{P}_n$ distribution becomes broader than a Poissonian.\\
As before, several simplifications are necessary.\\
This is the scenario that has been used in many of the experimental studies 
over the years. 

More recently a number of improvements have been included
\cite{petermi}.\\
1) The introduction of junction fragmentation, wherein the confinement
field between the three quarks in a baryon is described as a Y-shaped
topology, now allows the handling of topologies where several valence 
quarks are kicked out, thus allowing arbitrary flavours and showering
in all interactions in an event.\\
2) Parton densities are not only rescaled for energy--momentum 
conservation, but also to take into account the number of remaining
valence quarks, or that sea quarks have to occur in $q\overline{q}$ 
pairs.\\
3) The introduction of $p_{\perp}$-ordered showers allows the selection
of new ISR branchings and new interactions to be interleaved in one 
common sequence of falling $p_{\perp}$ values. Thereby the competition
between these two components, which both remove energy from the 
incoming beams, is modelled more realistically. FSR is not yet 
interleaved, but also does not compete for beam energy.\\ 
This scenario is not yet as well studied experimentally.

The traditional \textsc{Herwig} soft underlying event (SUE) approach to 
this issue has its origin in the UA5 Monte Carlo. In it a number of 
clusters are distributed almost independently in rapidity and transverse
momentum, but shifted so that energy--momentum is conserved, and the
clusters then decay isotropically. The multiplicity distribution of
clusters and their $y$ and $p_{\perp}$ spectra are tuned to give the 
observed inclusive hadron spectra. No jets are produced in this approach. 

The \textsc{Jimmy} \cite{jimmy} program is an add-on to \textsc{Herwig}. 
It replaces the SUE model with a MI-based one more similar to the 
\textsc{Pythia} ones above, e.g.\ with an impact-parameter-based picture 
for the multiple-interactions rate. Technical differences exist, e.g.\ 
\textsc{Jimmy} interactions are not picked to be $p_{\perp}$-ordered 
and thus energy--momentum issues are handled differently. 

The \textsc{DPMjet/DTUjet/PhoJet} family of programs \cite{phodtu} come 
from the ``historical'' tradition of soft physics, 
wherein multiple $p_{\perp} \approx 0$ ``pomeron'' exchanges fill a role 
somewhat similar to the hard MI above. Jet physics was originally not 
included, but later both hard and soft interactions have been allowed. 
One strong point is that this framework also allows diffractive events 
to be included as part of the same basic machinery. 

\subsection{Multiple-interactions studies}

How do we know that MI exist? The key problem is that it is not possible 
to identify jets coming from $p_{\perp} \approx 2$~GeV partons. Therefore 
we either have to use indirect signals for the presence of interactions 
at this scale or we have to content ourselves with studying the small 
fraction of events where two interactions occur at visibly large 
$p_{\perp}$ values.

An example of the former is the total charged multiplicity 
distribution in high-energy $pp/p\overline{p}$ collision. This 
distribution is very broad, and is even getting broader with
increasing energy, meaured in terms of the width over the average,
$\sigma(n_{\mathrm{ch}}) / \langle n_{\mathrm{ch}} \rangle$.
By contrast, recall that for a Poissonian this quantity scales like
$1/\sqrt{n_{\mathrm{ch}}}$ and thus is getting narrower. Simple models, 
with at most one interaction and with a fragmentation framework in 
agreement with LEP data, cannot explain this: they are way to narrow, and
have the wrong energy behaviour. If MI are included the additional 
variability in the number of interactions per event offers the missing 
piece \cite{maria}. The variable-impact-parameter improves the description 
further.

Another related example is forward--backward correlations. Consider the
charged multiplicity $n_f$ and $n_b$ in a forward and a backward rapidity
bin, each of width one unit, separated by a central rapidity gap of size
$\Delta y$. It is not unnatural that $n_f$ and $n_b$ are somewhat
correlated in two-jet events, and for small $\Delta y$ one may also be
sensitive to the tails of jets. But the correlation coefficient,
although falling with $\Delta y$, still is appreciable even out to
$\Delta y = 5$, and here again traditional one-interaction models    
come nowhere near. In a MI scenario each interaction provides additional 
particle production over a large rapidity range, and this additional 
number-of-MI variability  leads to good agreement with data.

Direct evidence comes from the study of four-jet events. These can 
be caused by two separate interactions, but also by a single one
where higher orders (call it ME or PS) has allowed two additional
branchings in a basic two-jet topology. Fortunately the kinematics
should be different. Assume the four jets are ordered in $p_{\perp}$,
$p_{\perp 1} > p_{\perp 2} > p_{\perp 3} > p_{\perp 4}$. If coming
from two separate interactions the jets should pair up into two
separately balancing sets, 
$|\mathbf{p}_{\perp 1} + \mathbf{p}_{\perp 2}| \approx 0$ and
$|\mathbf{p}_{\perp 3} + \mathbf{p}_{\perp 4}| \approx 0$. If an 
azimuthal angle $\varphi$ is introduced between the two jet axes
this also should be flat if the interactions are uncorrelated. 
By contast the higher-order graph offers no reason why the jets should 
occur in balanced pairs, and the $\varphi$ distribution ought to be
peaked at small values, corresponding to the familiar collinear
singularity. The first to observe an MI signal this way was the AFS 
collaboration \cite{afs} at ISR ($pp$ at 62 GeV), but with large
uncertainties. A more convincing study was made by CDF \cite{cdffourjet},
who obtained a clear signal in a sample with three jets plus a photon. 
In fact the deduced rate was almost a factor of three higher than naive 
expectations, but quite in agreement with the impact-parameter-dependent 
picture, wherein correlations of this kind are enhanced.

A topic that has quite extensively studied in CDF is that of the 
jet pedestal \cite{cdffield}, i.e.\ the increased activity seen 
in events with a jet, even away from the jet itself, and away from 
the recoiling jet that should be there. Some effects come from
the showering activity, i.e.\ the presence of additional softer
jets, but much of it rather finds its explanation in MI, as a kind of
``trigger bias'' effect, as follows. (1) Central collisions tend to 
produce many interactions, peripheral ones few. (2) If an event has 
$n$ interactions there are $n$ chances that one of them is hard. 
Combine the two and one concludes that events with hard jets are 
biased towards central collisions and many additional interactions.
The rise of the pedestal with triggger-jet energy saturates once
$\sigma_{\mathrm{int}}(p_{\perp \mathrm{min}} = p_{\perp \mathrm{jet}}) 
\ll \sigma_{\mathrm{nd}}$, however, because by then events are already 
maximally biased towards small impact parameter. And this is indeed
what is observed in the data: a rapid rise of the pedestal up to
$p_{\perp \mathrm{jet}} \approx 10$~GeV, and then a slower increase
that is mainly explained by showering contributions.

In more detailed studies of this kind of pedestal effects there are 
also some indications of a jet substructure in the pedestal, i.e.\
that indeed the pedestal is associated with the production of
additional (soft) jet pairs. 

In spite of many qualitative successes, and even some quantitative
ones, one should not be lead to believe that all is understood. 
Possibly the most troublesome issue is how colours are hooked up between
all the outgoing partons that come from several different interactions.
A first, already difficult, question is how colours are correlated
between all the partons that are taken out from an incoming hadron.
These colours are then mixed up by the respective scattering, in 
principle (approximately) calculable. But, finally, all the outgoing 
partons will radiate further and overlap with each other on the way out,
and how much that may mess up colours is an open question. 

A sensitive quantity is $\langle p_{\perp} \rangle (n_{\mathrm{ch}})$,
i.e.\ how the average transverse momentum of charged particles varies
as a function of their multiplicity. If interactions are uncorrelated
in colour this curve tends to be flat: each further interaction adds
about as much $p_{\perp}$ as $n_{\mathrm{ch}}$. If colours somehow
would rearrange themselves, so that the confinement colour fields would 
not have to run criss-cross in the event, then the multiplicity would 
not rise as fast for each further interaction, and so a positive slope
would result. The embarrassing part is that the CDF tunes tend to come
up with values that are about 90\% on the way to being maximally
rearranged \cite{cdffield}, which is way more than one would have 
guessed. Obviously further modelling and tests are necessary here 
\cite{peterrearr}.

Another issue is whether the $p_{\perp 0}$ regularization scale should 
be energy-dependent. In olden days there was no need for this, but
it became necessary when HERA data showed that parton densities rise
faster at small $x$ values than had commonly been assumed. This means 
that the partons become more close-packed and the colour screening 
increases faster with increasing collision energy.
Therefore an energy-dependent $p_{\perp 0}$ is not unreasonable, but
also cannot be predicted. Currently the default \textsc{Pythia} ansatz 
is $p_{\perp 0}(E_{\mathrm{cm}}) = (2.0~\mathrm{GeV}) \, 
(E_{\mathrm{cm}} / 1.8~\mathrm{TeV})^{0.16}$, i.e\ a predicted 
$p_{\perp 0} = 2.8$~GeV at 14 TeV. This gives a minimum-bias
multiplicity of about 7 per unit of rapidity in the central region,
and a pedestal under jet events of around 30 charged particles per unit
\cite{arthur}. However, these numbers are model- and parameter-dependent. 
\textsc{PhoJet} predicts about half as big a pedestal, and typical 
\textsc{Jimmy} tunes about 50\% more, a priori leaving a big range of 
uncertainty to be resolved once LHC runs begin.

\section{Hadronization}

The physics mechanisms discussed so far are mainly being played out on 
the partonic level, while experimentalists observe hadrons. In between
exists the very important hadronization phase, where all the outgoing
partons end up confined inside hadrons of a typical 1 GeV mass scale.
This phase cannot (so far?) be described from first principles, but
has to involve some modelling. The main approaches in use today are
string fragmentation and cluster fragmentation. These are described in 
the lectures of G.~Ingelman \cite{ingelman}, so this section will be
very brief, only with a few comments.

Hadronization models start from some ideologically motivated principles, 
but then have to add ``cookbook recipes'' with free parameters to arrive 
at a complete picture of all the nitty gritty details. This should come 
as no surprise, given that there are hundereds of known hadron species to 
take into account, each with its mass, width, wavefunction, couplings, 
decay patterns and other properties that could influence the structure 
of the observable hadronic state, and with many of those properties 
being poorly or not at all known, either from theory (lattice QCD) or 
from experiment. In that sense, it is sometimes more surprising that
models can work as well as they do than that they fail to describe 
everything.

The simpler initial state at an $e^+ e^-$ collider, such as LEP, implies
that this is the logical place to tune the hadronization framework to 
data \cite{lepdata}, and thereafter those tunes can be applied to other 
studies. One such is the internal structure of jets in hadron collider,
where the pattern in many respects is surprisingly well described.

On the other hand, at the HERA $e^{\pm}p$ collider it has been observed 
that the relative amount of strange-particle production is only $2/3$ of 
that at LEP, and of (anti)baryons only $1/2$. This has no simple 
explanation within the string fragmentation model, so it acts as a useful
reminder that we still do not know as much as we should. Also other 
examples could be provided.

\section{Summary and outlook}

In these lectures we have followed the flow of generators roughly 
``inwards out'', i.e. from short-distance processes to long-distance ones. 
At the core lies the hard process, described by matrix elements. It is 
surrounded by initial- and final-state showers, that should be properly
matched to the hard process. Multiple parton--parton interactions can
occur, and the colour flow is tied up with the structure of beam remnants.
At longer timescales the partons turn into hadrons, many of which are
unstable and decay further. This basic pattern is likely to remain 
in the future, but many aspects will change. 

One such aspect, that stands a bit apart, is that of languages.
The traditional event generators, like \textsc{Pythia} and \textsc{Herwig},
have been developed in Fortran --- up until the end of the LEP era this 
was the main language in high-energy physics. But now the experimental
community has largely switched to C++ and decisions have been taken e.g.\
at CERN to discontinue Fortran altogether. The older generators are still
being used, hidden under C++ wrappers, but this can only be a temporary 
solution, for several reasons. One is that younger experimentalists often
need to look into the code of generators and tailor some parts to specific
needs of theirs, and if then the code is in an unknown language this will
not work. Another is that theory students who apply for non-academic
positions are much better off if their resum\'es say ``expert on 
object-oriented programming'' rather than ``Fortran fan''.

A conversion program thus has begun on many fronts. \textsc{Sherpa},
as the youngest of the general-purpose generators, was conceived from the 
onset as a C++ package and thus is some steps ahead of the other programs 
in this respect. \textsc{Herwig++} \cite{herwigpp} is a complete 
reimplementation of the \textsc{Herwig}, as is \textsc{Pythia~8} of the 
current \textsc{Pythia~6}. Both conversions have taken longer than
originally hoped, but progress is being made and first versions exist.
\textsc{ThePEG} \cite{thepeg} is a generic toolkit for event generators, 
used by \textsc{Herwig++} and the upcoming new \textsc{Ariadne}. 
 
There are also other aspects where we have seen progress in recent years
and can hope for more:
\begin{itemize}
\item Faster, better and more user-friendly general-purpose matrix-element
generators with an improved sampling of phase space.
\item New ready-made libraries of physics processes, in particular with
full NLO corrections included.
\item More precise parton showers.
\item Better matching between matrix elements and parton showers.
\item Improved models for minimum-bias physics and underlying events.
\item Some upgrades of hadronization models and decay descriptions.
\end{itemize}
In general one would say that generators are getting better all the time,
but at the same time the experimental demands are also getting higher,
so it is a tight race. However, given that typical hadronic final states
at LHC will contain hundreds of particles and quite complex patterns
buried in that, it is difficult to see that there are any alternatives.

As the same time as you need to use generators, you should remain critical
and be on the lookout for bugs and bad modelling. Bjorken already many years 
ago worried about the passive attitude many experimentalists have towards 
the output of generators; they ``carry the authority of data itself. They 
look like data and feel like data, and if one is not careful they are 
accepted as if they were data.'' \cite{bjorken}. Don't you fall into that 
trap!

\end{document}